# EVALUATION OF THE DEBYE TEMPERATURE FOR THE IRON CORES IN HUMAN LIVER FERRITIN AND ITS PHARMACEUTICAL ANALOGUE FERRUM LEK USING MÖSSBAUER SPECTROSCOPY

S.M. Dubiel[1,✉], J. Cieślak[1], I.V. Alenkina[2,3], M.I. Oshtrakh[2,3,✉], V.A. Semionkin[2,3]

[1]*AGH University of Science & Technology, Faculty of Physics & Applied Computer Science, PL-30-059 Kraków, Poland*;
[2]*Department of Physical Techniques and Devices for Quality Control and*
[3]*Department of Experimental Physics,*
*Institute of Physics and Technology, Ural Federal University,*
*Ekaterinburg, 620002, Russian Federation*

**ABSTRACT**

An iron-polymaltose complex Ferrum Lek used as antianemic drug and considered as a ferritin analogue and human liver ferritin were investigated in the temperature range of 295–90 K by means of $^{57}$Fe Mössbauer spectroscopy with a high velocity resolution (in 4096 channels). The Debye temperatures $\Theta_D$=502±24 K for Ferrum Lek and $\Theta_D$=461±16 K for human liver ferritin were determined from the temperature dependence of the center shift obtained using two different fitting procedures.

✉ Corresponding authors: Stanislaw.Dubiel@fis.agh.edu.pl (S.M. Dubiel);
oshtrakh@gmail.com (M.I. Oshtrakh)

# 1. INTRODUCTION

Iron is essential for almost all living organisms by participating in a wide variety of metabolic processes like oxygen transport, DNA synthesis, and electron transport. However, iron concentrations in body tissues must be strictly regulated because excessive iron leads to tissue damage, as a result of formation of free radicals and iron overload. Disorders of iron metabolism are among the most common diseases of humans and encompass a broad spectrum of diseases with diverse clinical manifestations, ranging from iron deficiency anemia to iron overload and, likely, to neurodegenerative diseases [1–3]. In healthy organism the problem of the regulation of the iron concentration has been solved by storing iron in the iron storage protein ferritin that consists of 24 protein subunits shell and nanosized ferrihydrite ($5Fe_2O_3·9H_2O$) core [4]. The iron deficiency is a serious disease leading to anemia [1, 2]. In the case of the latter iron containing medicaments are used for treatment the iron deficiency. Some of them are manufactured in form of nanosized β–FeOOH cores surrounded with polysaccharide shells. These medicaments are often regarded as pharmaceutical analogues or model compounds of ferritin. The main differences between ferritin and its analogues are in the form of the iron core which is hexagonal ferrihydrite in ferritin and tetragonal β-FeOOH in the majority of analogues such as iron-dextran [5, 6] as well as in the shell. Comparative studies of ferritin and its analogues were done using various techniques such as extended X-ray absorption fine structure [7–9], electron diffraction [10], magnetization measurements [11–16] and Mössbauer spectroscopy (for reviews see [16, 17] and see, for instance, also [9, 18–28]). These studies demonstrated some similarities and differences of physical parameters for ferritin and its analogues. However, the main similarities in magnetization behavior and superparamagnetic behavior of Mössbauer spectra permitted to consider iron-polysaccharide complexes such as iron-dextran and iron-polymaltose complexes as ferritin analogues. The structure of the iron core in ferritin is intensively studied by various structural techniques such as electron nanodiffraction and high resolution electron microscopy. As a result, various models of the core – from monocrystalline to polyphasic – have been suggested [29–33]. It should be noted that Mössbauer spectra of ferritin and its analogues, both in paramagnetic and magnetic states, were also fitted using different approaches: 1) using a model-free way with a distribution function of quadrupole splitting or magnetic hyperfine field, 2) using one quadrupole doublet or one magnetic sextet, 3) using more than one spectral component. In the last-named case application of two components only (see [16]) was based on the core-shell model for the iron core suggested in magnetization study of ferritin [34]. Recently, using Mössbauer spectroscopy with a high velocity resolution we demonstrated small differences in hyperfine parameters for the iron cores in ferritins and its analogues obtained using one and several spectral components fits [26–28]. We have suggested considering one spectral component fit as the first (rough) approximation related to the homogeneous iron core model while several spectral components fit was related to the heterogeneous iron core model. To shed more light on the question concerning the similarity between the iron cores in ferritin and its analogues, a present Mössbauer spectroscopic study aimed at determining the Debye temperature of commercial iron-polymaltose complex Ferrum Lek and human liver ferritin was undertaken. The Debye temperature determined with Mössbauer spectroscopy, as isotope selective method, gives information only on iron atoms vibrations, hence it can be regarded as a relevant quantity to study possible structural differences between iron environment in the iron cores of ferritin and its synthetic approximant.

## 2. EXPERIMENTAL

A commercially available iron-polymaltose complex Ferrum Lek (Lek, Slovenia) in form of tablets was used as the ferritin analogue. For Mössbauer spectroscopic measurements 1/3 of a tablet was powdered by attrition. A lyophilized human liver ferritin with about 20% of bound iron was obtained from the Russian State Medical University, Moscow, Russian Federation (a process of the ferritin preparation was described elsewhere [23]). For the present study ferritin was used as powder with a sample weight of 100 mg. Both powders were placed in Plexiglas sample holders with a diameter of 20 mm and a height of 5 mm, and were pressed with a Plexiglas cover to exclude particles vibrations. The surface weights of the ferritin and Ferrum Lek samples for the Mössbauer measurements were about 5–6 mg Fe/cm$^2$ and 10 mg Fe/cm$^2$, respectively.

Mössbauer spectra were measured at the Ural Federal University (Ekaterinburg) using an automated precision Mössbauer spectrometric system built on the basis of the SM-2201 spectrometer with a saw-tooth shape velocity reference signal formed by a digital-analog convertor using quantification with 4096 steps (the high velocity resolution mode) and the temperature variable liquid nitrogen cryostat with moving absorber. Details and characteristics of this set-up are given elsewhere [35–37]. A $^{57}$Co in rhodium matrix source of about 1 GBq activity (Ritverc GmbH, Saint-Petersburg) kept at room temperature was used as a source of 14.4 keV gamma rays. The spectra of both samples were measured in transmission geometry with moving absorber kept in the liquid nitrogen cryostat at various temperatures in the range of 295–90 K and registered in 4096 channels.

Analysis of the measured spectra aimed at determining a center shift, *CS*, a spectral parameter of merit for determining the Debye temperature, $\Theta_D$. The spectra were analyzed with two different fitting procedures (A and B). According to A, each spectrum was analyzed in terms of one quadrupole doublet, as the first approximation, within the homogeneous iron core model with the following free parameters: the quadrupole splitting, $QS$, the center shift, $CS$, the line width, $\Gamma$, and the line intensity. The shape of the absorption line was assumed to be Lorentzian because the used Mössbauer spectrometric system demonstrated a pure Lorentzian line shape for the spectra of the standard absorbers (see [37]). The procedure B, as a model-independent fit, was based on an assumption that there is a distribution of $QS$, treated as a superposition of $N$ Gaussians. The distribution was assumed to be correlated with $CS$ (a parabolic correlation function was adopted). The number of the Guassians, $N$, was variable. The minimum of $\chi^2$-value was obtained for $N=7$ in the case of Ferrum Lek, and for $N=8$, in the case of ferritin. These distributions of $QS$ cannot be easily related to any regions/layers in the iron cores of human liver ferritin and Ferrum Lek like it was done for the fit using individual components in [28]. Therefore, the procedure B, despite giving evidence that the core is not homogeneous, can be considered as the first approximation also, and the average values of $CS$ obtained therefrom were used for a further determination of the Debye temperature. By doing so, we can compare values of the Debye temperature determined using both A and B procedures.

## 3. RESULTS AND DISCUSSION

Examples of the Mössbauer spectra of Ferrum Lek and human liver ferritin samples measured at 140 K with a high velocity resolution and fitted using procedures A and B are shown in Figs. 1–4. It is clearly seen that the fits within the rough approximation (procedure A) were not statistically satisfactory in comparison with those obtained using the procedure B.

The Debye temperature can be evaluated from Mössbauer-effect measurements in two independent ways i.e. from the temperature dependence of (1) $CS$, or (2) the recoil-free fraction, $f$ (related to spectral area). However, the two spectral quantities are related to different vibration properties of the lattice viz. $CS$ is related to the average squared velocity of vibrations, while $f$ is related to the average squared amplitude of vibrations. In other words they probe different aspects of the lattice dynamics, and, consequently, the values of $\Theta_D$ derived from them are, in general, different. For example for metallic iron $\Theta_D=421\pm30$ K was found from $CS(T)$, while $\Theta_D=358\pm18$ K was determined from $f(T)$ [38]. As $CS$ can be, in general, determined with a higher precision than $f$, consequently the accuracy of evaluation $\Theta_D$ from $CS$ is more reliable than that from $f$. We will, therefore, use the way (1) here for determining $\Theta_D$. The temperature dependence of *CS* can be expressed as follows:

$$CS(T) = IS(T) + SOD(T), \qquad (1)$$

where *IS* is the isomer shift and *SOD* is the so-called the second-order Doppler shift. Assuming that in the first-order approximation the phonon spectrum can be described by the Debye model, and taking into account that the temperature dependence of *IS* is weak, hence it can be neglected [39, 40], the temperature dependence of *CS* can be related to the Debye temperature via the second term in eq. (2):

$$CS(T) = -\frac{3kT}{2mc}\left(\frac{3\Theta_D}{8T} + 3\left(\frac{T}{\Theta_D}\right)^3 \int_0^{\Theta_D/T} \frac{x^3}{e^x - 1} dx\right), \qquad (2)$$

where $m$ is the mass of Fe atom, $k$ is the Boltzmann constant, $c$ is the speed of light, and $x = \hbar\omega/kT$ ($\omega$ being frequency of vibrations).

The temperature dependences of the *CS*-values as found from the Ferrum Lek Mössbauer spectra using the procedures described above and their best-fits in terms of equation (2) are illustrated in Fig. 5. The corresponding values of Debye temperature found thereby are as follows: $\Theta_D$=502±24 K from the approach A, $\Theta_D$=517±24 K from the approach B with 7 Gaussians and $\Theta_D$=518±26 K from the approach B with 40 Gaussians. It is clearly seen that the values of *CS* obtained from both A and B procedures applied to analyze the spectra are the same within the instrumental (systematic) error. Consequently, the values of $\Theta_D$ obtained from the *CS* temperature dependences were the same within the calculated error. The temperature dependences of the *CS*-values as found from the ferritin Mössbauer spectra using the procedures A and B were also the same within the instrumental (systematic) error. The values of Debye temperature determined from the *CS* temperature dependences using equation (2) were $\Theta_D$=461±16 K for the procedure A and $\Theta_D$=472±17 K for the procedure B. Again, as in the case of Ferrum Lek, both fitting procedures A and B gave the same value of $\Theta_D$ within the calculated error.

On the other hand, a very small difference in the values of $\Theta_D$ for Ferrum Lek and human liver ferritin obtained within the homogeneous iron core model as the first approximation gives evidence that average vibrations of iron atoms in the core of Ferrum Lek are approximately the same or slightly larger than those of the iron atoms in the ferritin core. The core structure of the former can be termed as slightly more rigid as the latter, possibly due to a higher degree of crystallinity and/or larger average size of the core (see [28]). We should also point out that the value of $\Theta_D$ for Ferrum Lek, an iron-polymaltose complex, obtained in the present study appeared to be significantly larger than $\Theta_D$=270±20 K found from the *CS(T)* measured in the temperature interval of 295–4 K for the polysaccharide iron complexes (Niferex) in [22]. In spite of the same form of the iron core (β-FeOOH) in both Ferrum Lek and Niferex (for the latter see [21]), these pharmaceutical products were produced by different manufactures in different conditions which, unfortunately, are unknown. Basing on the Mössbauer spectra of Niferex reported in [21], we can conclude that the anisotropy energy barrier for Niferex was higher than those for Ferrum Lek because in the 77 K Mössbauer spectrum of Niferex there was coexistence of paramagnetic doublet and magnetic sextet while in the 90 and 80 K Mössbauer spectra of Ferrum Lek there was a paramagnetic doublet only (see [28]).

## 4. CONCLUSIONS

Evaluation of the Debye temperature on the basis of the second-order Doppler shift in the Mössbauer spectra of human liver ferritin and its pharmaceutical analogue Ferrum Lek, an iron-polymaltose complex, measured with a high velocity resolution in the temperature range of 295–90 K was carried out. The Mössbauer spectra of the studied samples were fitted with two different procedures: A, using one quadrupole doublet to obtain the values of the *CS*, and B, using a distribution of quadrupole splitting to obtain the average values of *CS*. Both fits yielded the same values of *CS* that indicated that the one doublet fit was sufficient within the homogeneous iron core model as the first approximation of the Mössbauer spectra of ferritin and its analogues as far as the evaluation of the Debye temperature of the studied samples is concerned. Calculated values of the Debye temperature for the human liver ferritin and Ferrum Lek iron cores were 461±16 K and 502±24 K, respectively. The obtained differences are almost on the border of the errors limitation. Therefore, in the framework of this approach, the differences in the average iron atoms vibrations are very small. The latter indicates some similarities in the human liver ferritin and Ferrum Lek iron core structures related to iron dynamics. However, further development of the heterogeneous model to fit the Mössbauer spectra of ferritin and its analogues measured at various temperatures may permit us to evaluate the iron atoms vibrations in the iron core regions that have different spectral parameters.

**Acknowledgements**

This study was partly supported by the Ministry of Science and Higher Education, Warsaw and the Ministry of Education and Science of Russian Federation.

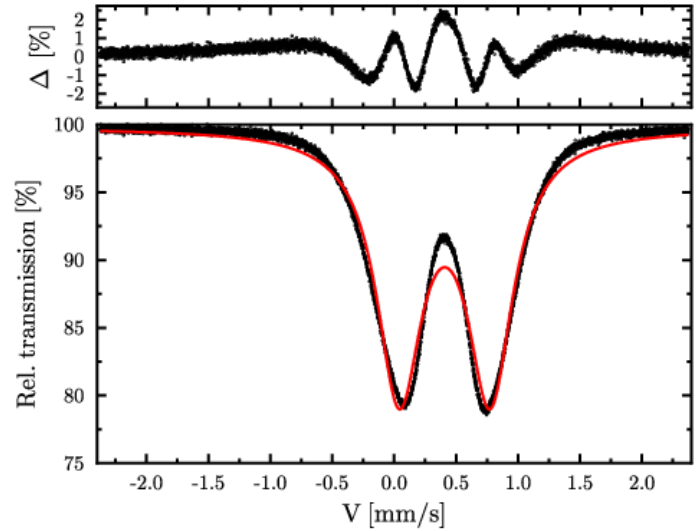

**Fig. 1.** $^{57}$Fe Mössbauer spectra recorded at 140 K on Ferrum Lek and fitted using one quadrupole doublet (procedure A). The best fit spectrum is shown as a solid (red) line. The difference spectrum is presented on top.

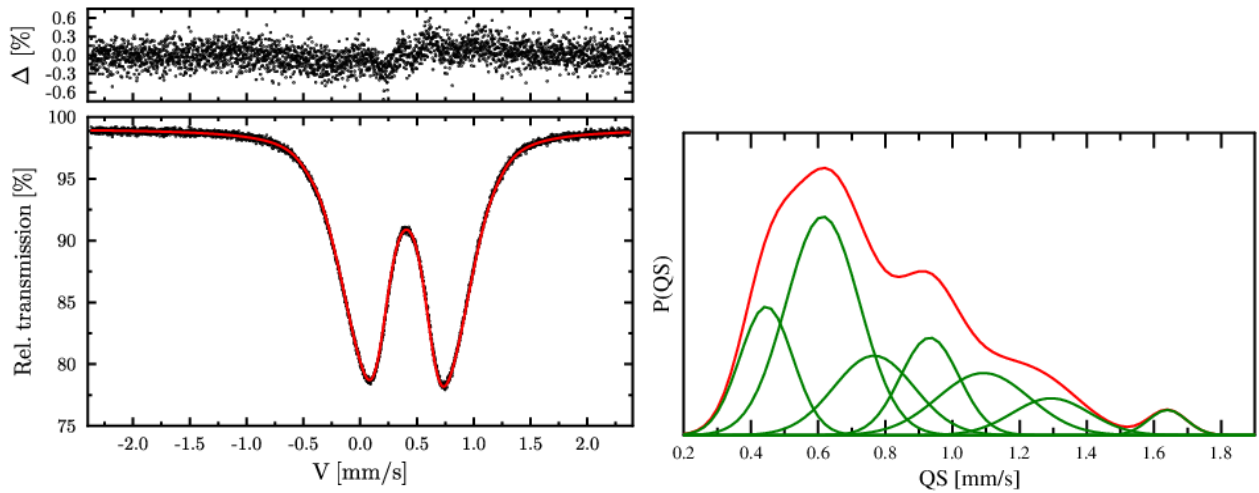

**Fig. 2.** (Left) $^{57}$Fe Mössbauer spectra recorded at 140 K on Ferrum Lek and fitted with a *QS*-distribution with 7 Gaussians (procedure B). The best fit spectrum is shown as a solid (red) line. The difference spectrum is presented on top. (Right) Distribution of *QS* as derived from the spectrum.

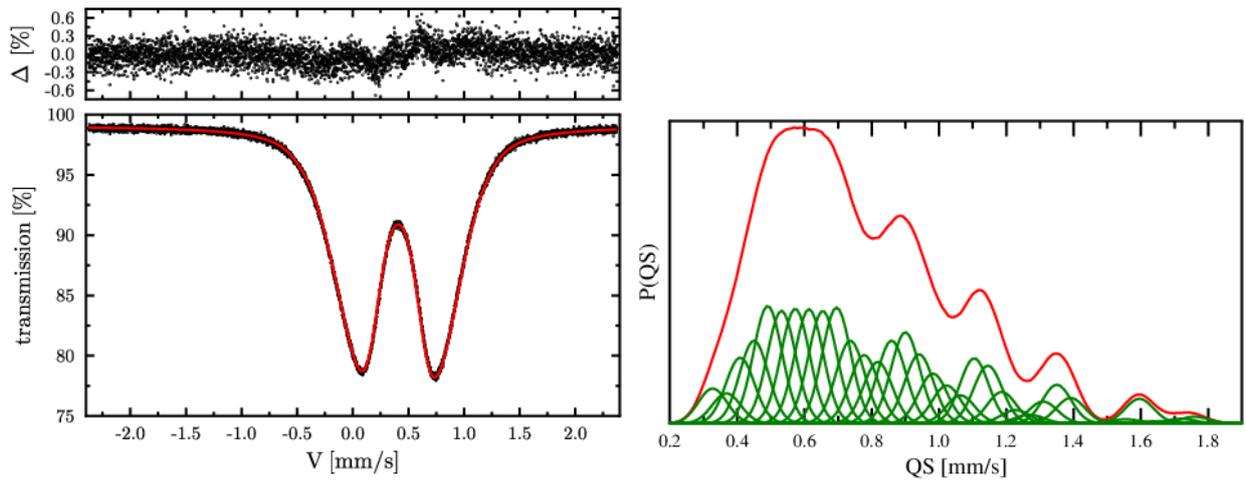

**Fig. 3.** (Left) $^{57}$Fe Mössbauer spectra recorded at 140 K on Ferrum Lek and fitted with a *QS*-distribution with 40 Gaussians (procedure B). The best fit spectrum is shown as a solid (red) line. The difference spectrum is presented on top. (Right) Distribution of *QS* as derived from the spectrum

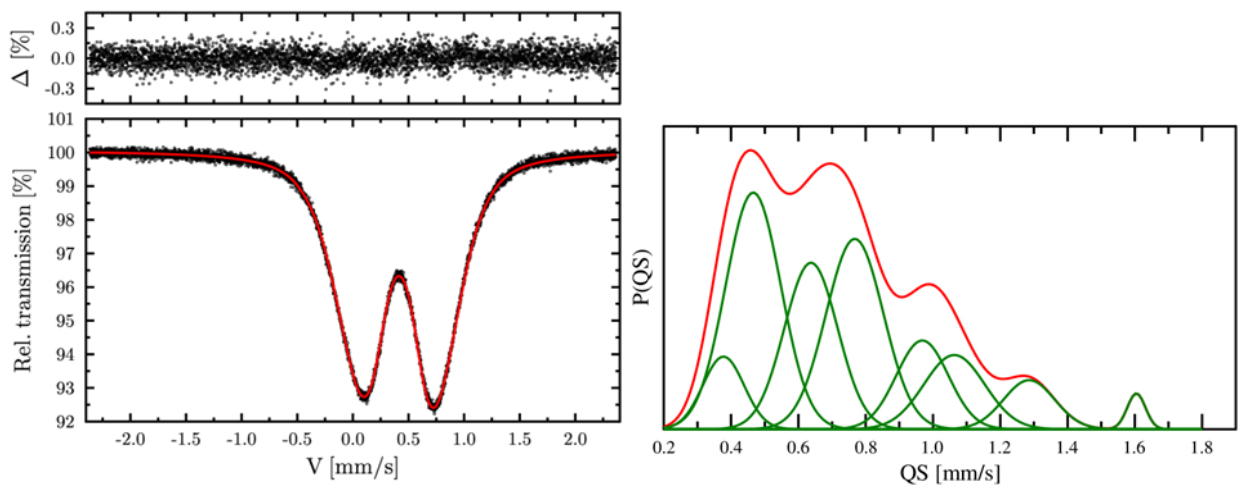

**Fig. 4.** (Left) $^{57}$Fe Mössbauer spectra recorded at 140 K on ferritin and fitted with a *QS*-distribution to 8 Gaussians (procedure B). The best fit spectrum is shown as a solid (red) line. The difference spectrum is presented on top. (Right) Distribution of *QS* as derived from the spectrum.

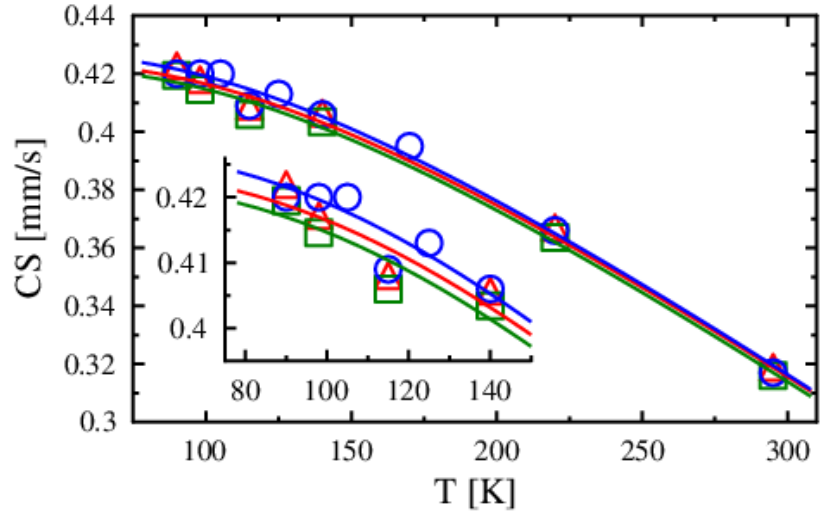

**Fig. 5.** Temperature dependences of the center shift, *CS* for the Ferrum Lek. Symbols represent the data obtained from the best-fit of the spectra to 1 quadrupole doublet (circles), to 7 Gaussians (triangles) and to 40 Gaussians (squares). Lines represent the best-fits to the data in terms of eq. (2). The inset shows the low temperature range.